\documentclass[floatfix,aps,twocolumn,prl,superscriptaddress, ]{revtex4}
\usepackage{amsmath}
\usepackage{amssymb}
\usepackage{graphicx}
\usepackage{dcolumn}
\usepackage{natbib}
\usepackage{bm}
\usepackage{verbatim}

\begin{document}

\title{Electron-phonon coupling in DyFeO$_3$ revealed by infrared spectroscopy}
\author{A. D. LaForge}
\email{andrew_laforge@cymer.com}
\affiliation{Departments of Physics and Electrical Engineering, University of California, Santa Cruz, Santa Cruz,
California 95064, USA}
\altaffiliation{Present address: Cymer, Inc., 17075 Thornmint Court, San Diego,
California 92127, USA}
\author{J. Whalen}
\affiliation{Department of Chemical and Biomedical Engineering, Florida State University, Tallahassee, FL 32306, USA}
\author{T. Siegrist}
\affiliation{Department of Chemical and Biomedical Engineering, Florida State University, Tallahassee, FL 32306, USA}
\author{A. P. Ramirez}
\affiliation{Departments of Physics and Electrical Engineering, University of California, Santa Cruz, Santa Cruz,
California 95064, USA}
\author{Z. Schlesinger}
\affiliation{Department of Physics, University of California, Santa Cruz, Santa Cruz,
California 95064, USA}
\date{\today }

\begin{abstract}
We have investigated crystal field and phonon dynamics of the multiferroic orthoferrite DyFeO$_3$ via polarized infrared spectroscopy. Reflectance of single crystals was  measured in the far- to mid-infrared spectral range at  range of temperatures from 10-295 K. We observe a strongly anisotropic phonon spectrum which differs from earlier lattice dynamical calculations in its symmetry, as well as a mode with significant and unusual temperature dependence that we interpret as a coupled phonon-crystal-field excitation.   
\end{abstract}

\maketitle

Multiferroic materials have long been of interest due to the rich variety of magnetically- and electrically-ordered phases they exhibit, and serve as excellent model systems for a range of spin-charge coupling phenomena. Further interest arises from the prospect of device applications which exploit magnetoelectric coupling to exert electric control over magnetic states, and vice-versa. Realization of such applications depends firstly on the energy scales associated with the ordered states; they must be large enough for macroscopic polarization effects to be robust at operational temperatures. Of the materials demonstrated to exhibit this quality, BiFeO$_3$ has been the most studied. Its ordering temperatures $T_{Neel}$ = 640 K \cite{Moreau-BiFeO3-Neel} and $T_{Curie}$ =  1090 K \cite{Lobo-BiFeO3} are remarkably high with respect to room temperature. The other criterion for device functionality, however, is strong magnetoelectric coupling, lattice-mediated interactions  between the spin and charge polarizations which permit their mutual control \cite{Cheong-multiferroic-review-NatMat2007}. This coupling is found in magnetically induced multiferroics such as CuO \cite{Kimura-Ramirez-CuO-induced-NatPhys2008}. Since BiFeO$_3$, an improper ferroelectric, derives its polarization from a lone-pair dipole phenomenon, and not from distortions in the ionic lattice, it does not possess the necessary coupling for applications, leading to a search for new materials that exhibit stronger magnetoelectric coupling. 

Recently, attention has fallen to the rare-earth orthoferrite (RFeO$_3$) series \cite{White-OF-review-JAP1969}, whose orthorhomically-distorted perovskite structure contains two magnetic subsystems and tilted oxygen octohedra, providiving ample opportunities for magnetoelectric coupling via the Dzayloshinskii-Moriya interaction \cite{Dzaylo-Moriya}. SmFeO$_3$ was recently shown to be an improper ferroelectric, with ME coupling over a range of temperatures \cite{Lee-SmFeO3-ferroelctric-PRL2011}. DyFeO$_3$ has been demonstrated \cite{Tokunaga-Tokura-DyFeO3-PRL2008} to exhibit ME coupling, but only at temperatures below the rare-earth Neel temperature T$_N$ = 4 K, and in the presence of an applied magnetic field. A spin reorientation transition at 37 K, however, suggests that higher energy scales may be at play.

Given the importance of the lattice in mediating interactions between the magnetic and electric polarizations, a detailed understanding of the lattice vibrational modes and their couplings to spin and charge symmetries is essential to the advancement of multiferroic-based technologies. Infrared spectroscopy is an especially effective tool for exploring the lattice spectra; it offers high energy resolution and sensitivity for detecting infrared-active lattice vibrations, and reveals interactions between phonons and other dipolar excitations, both electric and magnetic. 

Here we report measurements of the reflectivity of DyFeO$_3$ in the infared frequency range, and utilize an multi-oscillator model fit to parameterize the observed absorptions. We find that the experimental eigenfrequencies and symmetries are not in agreement with those calculated in previous work on orthoferrites \cite{Gupta-lattice}. For electric fields along the orthorhombic [$\overline{1}$ 1 0] crystal axis we observe the low-temperature emergence of an absorption corresponding to a  crystal field level transition; this mode is coupled to the lowest frequency phonon, leading to notable and highly unusual shifts of eigenfrequency and linewidth. 

The single crystals of DyFeO$_3$ measured in this study were grown by a flux method by J. P. Remeika at Bell Laboratories. A 2x2x2.5 mm$^3$ sample was selected for magnetization measurements, while a much larger crystal from the same batch, with a mirror-like, flat natural face, was used for optical and x-ray experiments.  X-ray diffraction confirms the orthorhombic structure, with lattice constants $a$ =5.3 \AA, $b$=5.6 \AA, and $c$=7.6 \AA. Sharp diffraction peaks indicate that the sample is single-phase, without garnet inclusions or flux contamination. Magnetization was measured in a SQUID magnetometer (Quantum Design MPMS) with magnetic fields along the the principle pseudo-cubic axes.

Infrared reflectance spectra were recorded at near-normal incidence for frequencies $\omega$ = 50-5000 cm$^{-1}$ and temperatures $T$ = 10-300 K. Incident electric fields were polarized along either the [1 1 0] or [$\overline{1}$ 1 0] axes, corresponding to the unit axes of the high-temperature cubic unit cell, rather than those of the orthorhombic structure. (See inset to Fig. 1.) Since the tilting of oxygen octohedra is relatively small the crystal retains a pseudocubic symmetry. As we show below, these measurement axes provide the best basis for understanding the electrodynmic response.

Magnetic susceptibility $\chi_M$ is shown in Fig. 1(b) as a function of temperature for magnetic field $H$ = 1000 Oe applied parallel to the [1 1 0], [$\overline{1}$ 1 0], and [0 0 1] axes. Below $T_N^Fe$ the Fe ions are ordered antiferromagnetically in the [1 0 0] (G-type) and [0 1 0] (A-type) directions, with a small sublattice canting producing a weak ferromagnetic moment (WFM) in the [0 0 1] direction \cite{White-OF-review-JAP1969,Tokunaga-Tokura-DyFeO3-PRL2008}.  At the spin reorientation transition temperature $T_{SR}^Fe$ = 37 K the spin structure rotates from $G_x A_y F_z$ to $A_x G_y C_z$, and the WFM disappears. Dy ions order at $T_N^{Dy}$ = 4.5 K. 

\begin{figure}
\centering
    \includegraphics[width=3.375in]{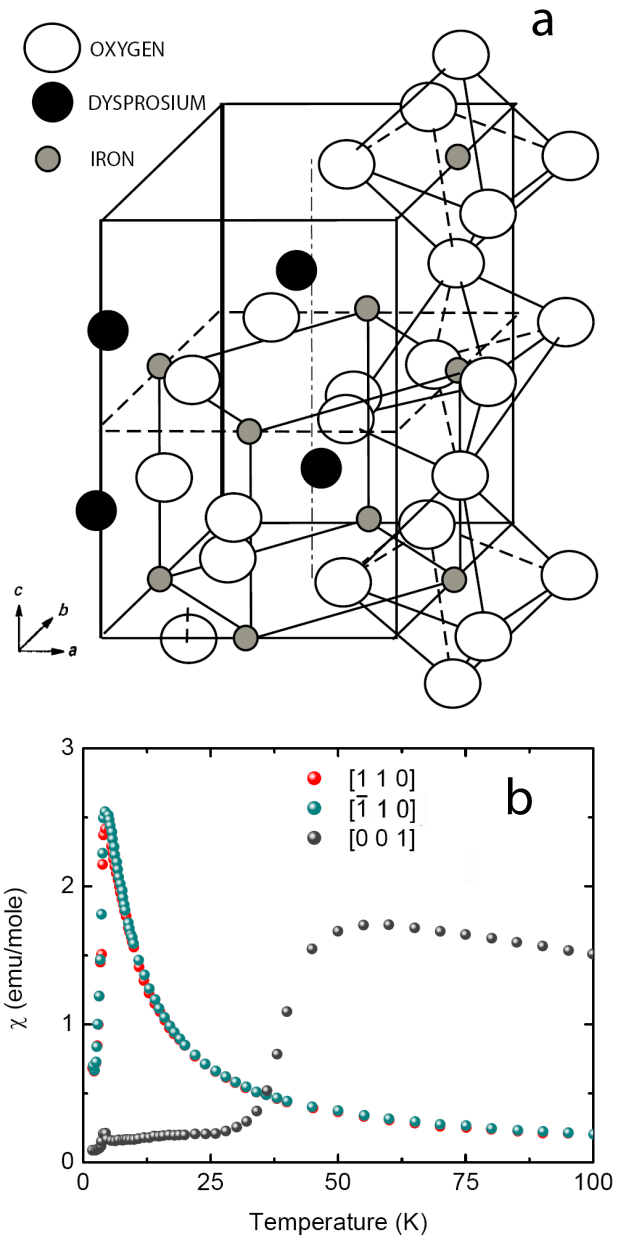}
     \caption{(a) Crystal structure for orthorhombically distorted perovskite DyFeO$_3$. Orthorhombic axes are rotated 45$^{\circ}$ relative to the cubic unit cell. (b) Magnetic susceptibility $M/H$, collected with $H = 1000$ G oriented along the principle pseudocubic axes. } 
     \label{Fig1}
\end{figure}

Reflectivity of DyFeO$_3$ in the far-infrared frequency range, shown in Fig. 2 for selected temperatures, is dominated by lattice vibrational modes.  These modes generally can be assigned to three different types of motion: Dy external modes at the lowest frequencies, Fe-O bending modes in the intermediate range, and oxygen stretching modes at higher frequencies. At room temperature the differences between spectra taken with [1 1 0] (Fig. 1a) and [1 -1 0] (Fig. 1b) polarizations are readily apparent, with 16 phonons observed in the former and 12 in the latter. As the temperature is lowered to $T$ = 10 K, all modes in the [1 1 0] spectra harden and narrow slightly due to nonlinear effects.  The [$\overline{1}$ 1 0] spectra, however, are strongly renormalized at low temperature. In particular, a new mode appears at 140 cm$^{-1}$ for $T \lesssim$180 K. 

Spectra recorded at these two polarizations form a linearly independent basis; that is, spectra measured at intermediate polarizer angles contain an admixture of the basis spectra. This strong association of the phonon spectra to the pseudocubic unit cell is at odds with the dynamical lattice calculations of Ref. \onlinecite{Gupta-lattice}, which predict phonon modes of type $B_{1u}$, $B_{2u}$, and $B_{3u}$ polarized along the orthorhombic crystal axes. This discrepancy makes phonon assignment difficult, and highlights a need for further effort in calculating the lattice dynamics. 

In order to quantify the electrodynamic response of this system we fit the reflectance spectra to a sum of Drude-Lorentz oscillators of  the form
\begin{equation} 
\epsilon_{\pm}(\omega)=\epsilon_\infty+\sum\limits_{n=1}^N \frac{\omega^2_{p,n}}{\omega^2_{0,n}-\omega^2-i \gamma_n \omega},
 \end{equation}
 where $\omega_{p,n}$ is the plasma frequncy, $\omega_{0,n}$ is the transverse resonance frequency, and $\gamma_n$ is the linewidth of the $n$th oscillator. A fit to the four-parameter product form of the Drude-Lorentz oscillator \cite{Barker-Hopfield-4osc-model}, which takes into account LO-TO phonon splitting, was also carried out; however, since the fit was not substantially improved, the more intuitive 3-parameter fit results will be discussed here. Best-fit eigenfrequencies and linewidths are shown in Fig. 3, scaled to their values at 180 K. As seen in Fig. 3a, temperaure-induced changes to [1 1 0] phonon modes are less than 2\%, and are qualitatively similar for all modes in the far infrared region. 

For $E \parallel$ [$\overline{1}$ 1 0], however, the lowest-frequency modes change dramatically with temperature. The lowest-frequency mode, labeled mode TO1, is only resolved below $T \approx$ 180 K, and exhibits an 7\% downshift in eigenfrequency upon cooling to 10 K. The  next highest mode, in contrast, hardens significantly as temperature decreases, with a 7\% increase in transverse frequency. 

The temperature dependence of phonon damping is also anomalous.  Unlike the case for [1 1 0], where all modes exhibit nearly identical narrowing of  20-50\%, for [$\overline{1}$ 1 0] a range of damping values are observed, with the greatest changes occurring in the lowest-frequency modes. For TO2, the  low-frequency mode that hardens as temperature is lowered,  $\gamma$(10 K) drops as low as 0.15$\gamma$(180 K).  The linewidth of TO1 is especially peculiar, 
increasing slightly after TO1 is first observed, then decreasing sharply below $T =$ 135 K.

\begin{figure}
\centering
    \includegraphics[width=3.375in]{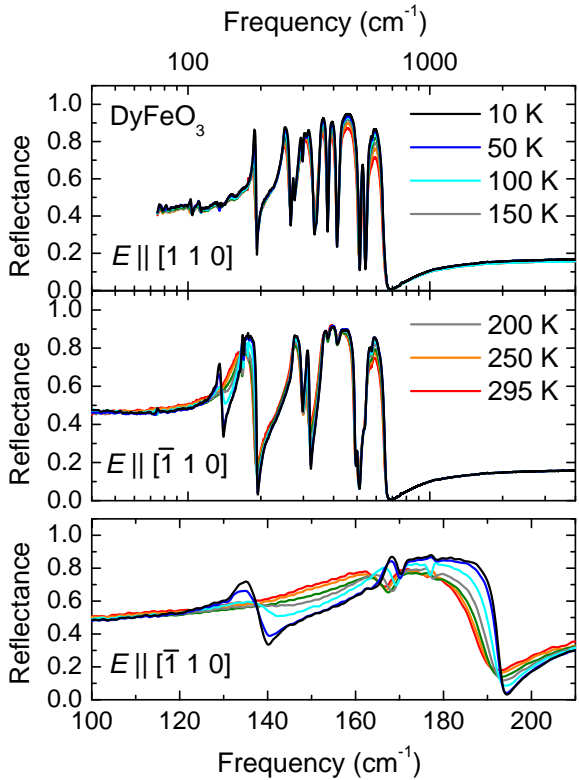}
     \caption{(a) Infrared reflectivity for DyFeO$_3$ with $E \parallel$ [1 1 0] polarization. (b), (c) Reflectivity for [$\overline{1}$ 1 0] polarization.} 
     \label{Fig2}
\end{figure}

\begin{figure}
\centering
    \includegraphics[width=3.375in]{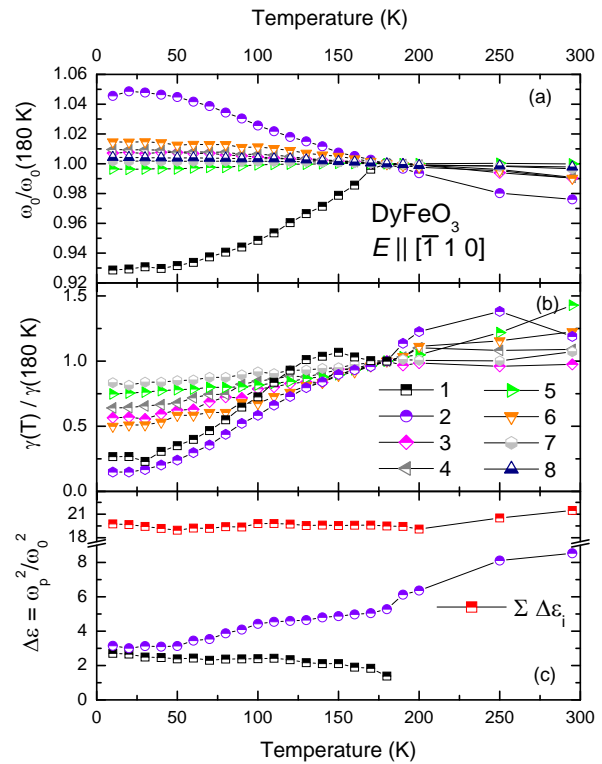}
     \caption{Multi-oscillator fit results for $E \parallel$ [$\overline{1}$ 1 0] polarization. Fit  parameters (a) eigenfrequency and (b) linewidth are scaled to their values at $T$= 180 K. (c) Dielectric strength for the two lowest frequency modes, as well as total dielectric strength from all infrared modes.} 
     \label{Fig3}
\end{figure}

In interpreting the origin of the TO1 mode observed here, recall that there are many electronic processes which produce far-infrared absroption signatures.  We now discuss several of these possibilities and conclude that the explanation most consistent with the data is that of a crystal field transition coupled to two nearby phonons. The first possibility is that TO1 is a lattice mode. Indeed, it has comparable frequency and oscillator strength to nearby phonon modes, but its sudden onset and narrowing at the intermediate temperature scale of 135-180 K  would be difficult to understand in the absence of a structural phase transition. Such a transition has not been previously reported in related rare-earth orthoferrites; thus, this assignment is highly unlikely. Dramatic decreases in phonon transverse frequency and oscillator strength are commonly seen in systems with spontaneous dielectric polarization, as lattice vibrations with eigenvectors corresponding to the polarization vector become unstable. However, a study of the dielectric constant in DyFeO$_3$ only found spontaneous polarization for $T<$ 4 K and magnetic fields greater than 24 kOe \cite{Tokunaga-Tokura-DyFeO3-PRL2008}. The present optical data agree with this result, as the Lydanne-Sachs-Teller sum rule dictates that such changes in the dielectric strength of finite-frequency oscillators must be offset by an increase in the low-frequency dielectric value $\epsilon_0$. As seen in Fig. 3(c), the sum of the dielectric strength of all observed oscillators remains constant, and $\epsilon_0$ is unchanged. In fact, most of the dielectric strength gained by the new mode TO1 is offset by a decrease in the strength of TO2. 

Given the high ferromagnetic ordering temperature and the antiferromagnetic and spin-reorientation transitions at low temperatures, it is worth considering whether the TO1 mode could be magnetic in origin. However, the large dielectric strength and relatively high frequency are inconsistent with an assignment of a magnetic dipole transition. Electrically-excited magnons (electromagnons) have been observed in the related perovksite manganite (RMnO$_3$) compounds; these excitations are most commonly observed at sub-THz frequencies and in regions of the temperature-magnetic field phase diagram with magnetically-induced ferroelectricity, but they have been found in non-multiferroic phases, and at higher frequencies and temperatures \cite{Takahashi-TbMnO3,Shuvaev-GdMnO3}. In the latter cases the excitations are often related to two-magnon processes, emerging out of spin fluctuations at temperatures severals times greater than the ordering temperature \cite{Takahashi-TbMnO3,Shuvaev-GdMnO3}.  In the case of DyFeO$_3$, however, the spin structure lacks noncollinearity, and thus does not produce the dielectric polarization and spin-lattice coupling central to an electromagnon. Also, the intermediate temperature scale at which TO1 turns on (180 K) is commensurate with neither the rare-earth $T_{Neel}$ (4.5 K) nor the Fe $T_C$ (745 K), and the mode is insensitive to the change in magnetic symmetry at $T_{SR} = 37$, further discounting the possibility of an origin in a collective magnetic phenomenon. 

\begin{figure}
\centering
    \includegraphics[width=3.375in]{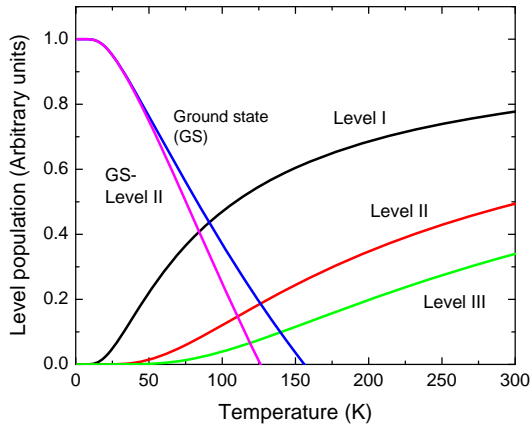}
     \caption{Crystal field level populations calculated using Boltzmann statistics} 
     \label{Fig4}
\end{figure}

We propose that the TO1 mode is the signature of an electric dipole transition between crystal field-split levels of the electronic groundstate. Earlier optical studies of DyFeO$_3$ \cite{Schuchert-Faulhaber-DyFeO3-ZPhys1969} have revealed $^6H_{15/2} \rightarrow ^6F_{5/2}$ transitions with crystal field levels I, II, and II at 52, 147, and 225 cm$^{-1}$, respectively. The second level (147 cm$^{-1}$) is close in energy to the TO1 mode, and matches its extrapolated value at 200 K (near the temperature where TO1 is first resolved). These levels are thermally populated  according to Boltzmann statistics, and their relative populations $P_i$ are have been calculated and plotted in Fig. 4 as a function of temperature. Transitions from the ground state to level $i$ begin when $P_0-P_i  > 0$; this onset is calculated to be 125 K for level II, in rough agreement with the data. 

We suggest that the anomalous temperature dependence and of both the TO1 and TO2 modes, as well as the transfer of spectral weight between them, represents the realization of a coupled crystal-field-phonon excitation \cite{Thalmeier-Fulde-CF-phonon-coupling-PRL1982,Wegerer-CF-phonon,Stach-CF-phonon}. Such a coupling can only be observed when a rare-earth crystal field (CF) transition lies close in energy to a phonon of the same symmetry, and the two resultant excitations of the composite system assume a mode-repulsion character, diverging in energy with increasing interaction strength. CF-phonon coupling is made allowed in this system due to the distorted crystalline electric field at the orthorhombic Fe sites.  Detailed calculations \cite{Thalmeier-Fulde-CF-phonon-coupling-PRL1982,Wegerer-CF-phonon,Stach-CF-phonon} based on a simple two-level model have shown that, for coupled two-peak structures in Raman scattering data, the unrenormalized mode positions and intensity ratio can be used to determine the bare CF excitation and phonon frequencies $\omega_{CF}$ and $\omega_{ph}$, as well as their interaction strength $V$. Here we extend this analysis by analogy to modes of infrared activity to extract $\omega_{CF}$, $\omega_{ph}$, and $V$ via

\begin{equation}
\omega_{CF}=\frac{\omega_{0,1}+R\,\omega_{0,2}}{1+R},
\label{omegaCF}
\end{equation}

\begin{equation}
\omega_{ph}=\frac{R\:\omega_{0,1}+\omega_{0,2}}{1+R},
\label{omegaph}
\end{equation}

and

\begin{equation}
V=|\omega_{0,1}-\omega_{0,2}|\frac{\sqrt{R}}{1+R}, 
\label{Vcoupling}
\end{equation}

where $R=\omega_{p,1}^2/\omega_{p,2}^2$. The bare values have been calculated from parameters obtained from the Lorentzian oscillator fits to the reflectivity and plotted in Fig. 5. The interaction strength $V$ increases as the temperature is lowered, reflecting the greater splitting between modes and relative gain of spectral weight by the TO1 mode. The bare frequencies, in contrast to the observed ones, are nearly constant with temperature, with $\omega_{CF}=147-149$ cm$^{-1}$ closely matching the value of 147 cm$^{-1}$ from earlier optical studies \cite{Schuchert-Faulhaber-DyFeO3-ZPhys1969}.

\begin{figure}
\centering
    \includegraphics[width=3.375in]{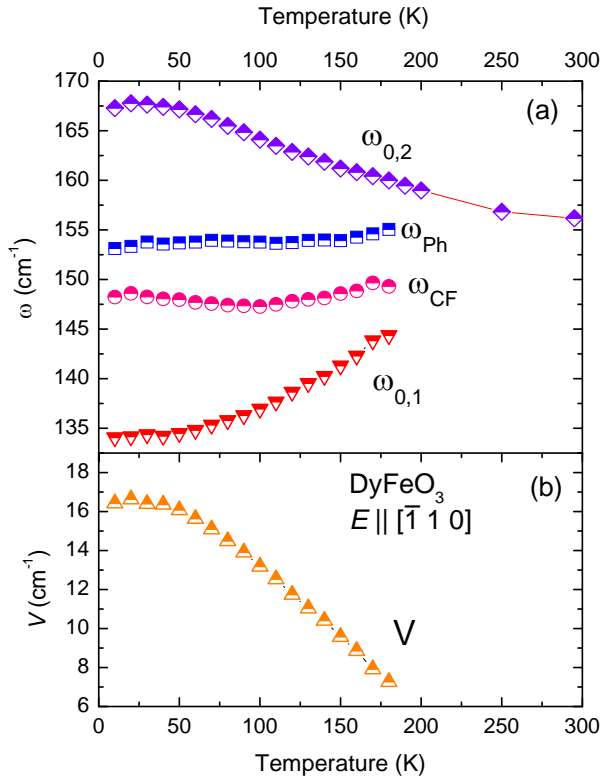}
     \caption{(a) Bare crystal field and phonon frequencies extracted from experimentally observed frequencies. (b) Crystal-field-phonon coupling parameter $V$. } 
     \label{Fig5}
\end{figure}

In summary, we have observed phonon-crystal-field coupling in the muliferroic perovskite DyFeO$_3$ via infrared spectroscopy. Calculation of the bare CF excitation and phonon energies yields excellent agreement with earlier optical meaurements, adding credence to this interpretation of the infrared data. These findings yield new insight into the nature of the interaction of the crystal lattice with electronic degrees of freedom, an essential ingredient in the realization of new multiferroic phenomena. More detailed lattice dynamical calculations will improve our understanding of the symmetries of the modes involved and their relation to the electronic and magnetic structure.

\section*{Acknowledgments}
ADL gratefully acknowledges discussions with Andrei Sushkov and Jan Musfeldt.  Research at UCSC was supported by NSF.  JW and TS acknowledge support from the Florida State University, the State of Florida, and the NHMFL (NSF Cooperative Agreement No. DMR-0654118, and the U.S. Department of Energy).

\end{document}